\documentclass[twocolumn,pra]{revtex4}
\usepackage{amssymb}
\usepackage{graphicx}
\usepackage{dcolumn}
\usepackage{bm}
\usepackage{amsmath}
\usepackage{epsfig}
\usepackage{braket}

\begin{document}

\title{Orienting  Asymmetric Molecules by Laser Fields with  Skewed Polarization}

\author{E. Gershnabel}
\author{I. Sh. Averbukh}
\affiliation{Department of Chemical Physics, The Weizmann Institute
of Science, Rehovot 76100, ISRAEL}
\begin{abstract}

We study interaction of generic asymmetric molecules with a pair of strong time-delayed short laser pulses with crossed linear polarizations. We show that such an excitation not only provides unidirectional rotation of the most polarizable molecular axis, but also induces a directed torque along this axis, which results in the transient orientation of the molecules. The asymmetric molecules are chiral in nature and different molecular enantiomers experience the orienting action in opposite directions causing out-of-phase oscillation of their dipole moments. The resulting microwave radiation was recently suggested to be used for analysis/discrimination of chiral molecular mixtures. We reveal the mechanism behind this laser induced orientation effect, show that it is classical in nature, and envision further  applications of light with skewed polarization.
\end{abstract}

\maketitle

\section{Introduction} \label{Introduction}
The orientation and alignment of molecules have long been of interest in chemistry and physics.
Applications of aligned and oriented molecules, such as high harmonic generation  \cite{Velotta}, chemical reactivity \cite{Warren}, control of photodissociation and photoionization processes \cite{Larsen},  and attosecond molecular dynamics \cite{Attosecond} have motivated the development of all-optical techniques for controlling molecular angular degrees of freedom.
A major advance has been the use of linearly polarized, ultrashort laser pulses to create aligned molecular states at field-free conditions by an impulsive Raman mechanism (for a review of laser molecular alignment, see Refs. \cite{Tamar,Ohshima10,Fleischer12,Krems}). Recent studies (both theoretical and experimental) of the alignment process have focused on the control of molecular deflection (see, e.g. \cite{bib40}, and references therein), physics of molecular superrotors \cite{Mullin,Korobenko2014,Superrotors1} and the analysis of molecular dissipation effects \cite{Superrotors1,Collisions1,Collisions3,Collisions2}.
Molecular orientation, on the other hand, provides additional features such as generation of even harmonics \cite{Kraus}, directional molecular ionization or dissociation \cite{De} and control of coherent radiative decay of molecular rotations \cite{FleischerDecay}. Orientation of molecules requires interactions that define a direction in space (distinguish between "up" and "down"),  and a variety of methods have been proposed to break the symmetry, i.e. by using a combination of laser and dc electrostatic fields \cite{Friedrich1,Sakai,Ghafur,Holmegaard,Goban}, multi-frequency ($\omega+2\omega$) fields \cite{De,Zheng2010,Frumker}, single-cycle THz pulses \cite{Harde,Fleischer1,Kitano,Machholm,Dion}, or their combination with the optical pulses \cite{Daems,Erez1,Jones,Fleischer2}.

In this work we describe and investigate a  mechanism in which orientation of generic asymmetric molecules is achieved by a  chiral skewing of the polarization of short non-resonant laser pulses. Fast oscillating linearly polarized optical fields hardly couple to the permanent molecular dipole moment, but interact with the induced molecular polarization in an axially symmetric way. Such fields define a non-directional polarization axis in space, about which the molecules tend to align, but cannot be oriented. If however the polarization axis turns with time in some plane, and this rotation has a certain sense - this defines a directed vector which is perpendicular to that plane and has  direction determined by the sense of rotation. The simplest example of such a field is a pair of  delayed laser pulses with crossed linear polarizations. Such an excitation was suggested in \cite{bib22,bib23} for inducing unidirectional molecular rotation (UDR). The resulting orientation of molecular angular momentum was experimentally demonstrated in \cite{bib23,Korech,Kenta,JW}, investigated in detail, both from the quantum and classical perspectives, in \cite{bib24}, generalized to chiral trains of multiple pulses in \cite{Valerytrain,Floss,Bloomquist} and to laser fields with continuously twisted polarization in \cite{Karras2015}.

In what follows, we consider asymmetric molecules with a permanent dipole moment and anisotropic polarizability. For a generic asymmetric molecule (which is lacking all symmetry elements except the trivial identity), the direction of the dipole moment is different from any of the principal molecular axes of inertia. In addition, the polarizability tensor is generally non-diagonal in the molecular frame, which means that an electric field applied along one principle axis induces dipole moments along the other two as well.  Moreover,  generic anisotropic molecules are inherently chiral \cite{Cotton}. We will demonstrate that interaction of the two chiral entities - an anisotropic molecule and laser field with skewed polarization - results in partial orientation of the isotropic ensemble of molecules and their dipole moments. The orientation direction depends on the mutual handedness of the molecules and the skewed polarization. The resulting transient macroscopic polarization of the gas can be measured by means of the free-induction decay signal emitted by the sample, similar to \cite{Harde,Bigourd,Fleischer1,Fleischer2,Faucher2016}. One  of the applications of this orientation effect relates to the analysis of chiral molecular mixtures in gas phase, as was recently theoretically suggested in \cite{bib33} based on quantum mechanical arguments. By considering excitation of a mixture of two molecular enantiomers  by a pair of laser pulses with crossed polarization, it was shown \cite{bib33} that the resulting emission from the gas  bears  information on the chiral composition of the molecular mixture. We show that the physics behind this approach is of purely classical origin, and we qualitatively describe the underlying orientation mechanism in Section II. In Section III, a comprehensive analysis of the orientation effect in an initially isotropic thermal ensemble of asymmetric molecules is provided by considering fully classical rotational dynamics of molecules subject to a pair of UDR laser pulses with skewed polarisation \cite{bib22,bib23}. The results are discussed in Section IV, and their connection to the enantiomers differentiation problem \cite{bib33} is outlined in Section V.   Finally, we summarize our findings and discuss further directions in Section VI.

\section{Qualitative Classical Analysis} \label{Simplified Classical Analysis}
The orientation scheme discussed in this work considers generic asymmetric  molecules with anisotropic polarizablity and a permanent dipole moment, which are subject to two linearly polarized time delayed short laser pulses that induce unidirectional rotation \cite{bib22,bib23,bib24}. We define the laboratory axes as $X,Y$ and $Z$. The first pulse is polarized along the $X$ direction and the second pulse is polarized in the $XY$ plane at $45^{\circ}$ to the $X$ axis. The molecules are initially at thermal equilibrium.

\begin{figure}[htb]
\begin{center}
\includegraphics[width=8cm]{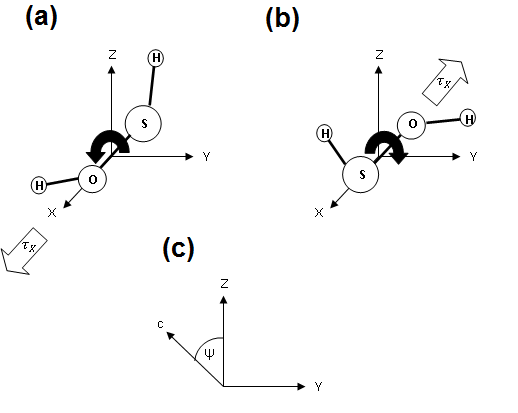}
\end{center}
\caption{Orientation mechanism. Just before the second pulse, the molecular $a$ axis  points either toward $+X$ direction (a), or to the $-X$ direction (b). The second pulse causes a torque directed either along the $+X$ direction (a) or the $-X$ direction (b). As a result, the molecules start rotating about the ${\rm O-S}$ bond, and in both cases the projection of the torque along the ${\rm O-S}$ bond is the same (as is illustrated by the black arrows). This leads to the transient orientation of the molecules. (c) Angle $\psi$ between the $c$ axis and the $Z$ axis in the $YZ$ plane  is uniformly distributed between $0^{\circ}$ and $360^{\circ}$ by the time of arrival of the second pulse.}
\label{Simplified figure}
\end{figure}

In order to illustrate the effect, we use the ${\rm HSOH}$ molecule previously considered in \cite{bib33}, and model it as a classical asymmetric rotor with principal axes $a,b$ and $c$ (the $a$ axis is along the ${\rm O-S}$ bond). Since the $a$ axis is the most polarizable one \cite{bib33}, the first laser pulse aligns it along the $X$ direction. For the sake of clarity,  we consider a simplified situation, where all the molecules in the ensemble are completely aligned along the $\pm X$ direction just before the second pulse, as  illustrated in Fig. \ref{Simplified figure}.
In other words, the direction of the $a$ axis (${\rm O-S}$ bond) is either along or against the $X$ axis, as can be seen in Figs. \ref{Simplified figure}a and \ref{Simplified figure}b, respectively. The angle $\psi$  shown in Fig. \ref{Simplified figure}c is the angle between the $c$ principal axis of the asymmetric rotor and the $Z$ axes in the $YZ$ plane, and it is uniformly distributed in the interval [$0^{\circ} , 360^{\circ}$].

When the second pulse is applied in the $XY$ plane,  the electric field,  $\overrightarrow{E}$ of the pulse polarizes the molecules, and induces the dipole moment:

\begin{equation}
d_i=\sum_j\alpha_{ij}E_j,
\label{Dipole}
\end{equation}
where $i,j=X,Y,Z$. The torque exerted on the molecule is given by:
\begin{equation}
\vec{\tau}=\vec{d}\times \vec{E}.
\label{Torque}
\end{equation}
In our case, the components of the electric field of the second pulse are $E_X=E_Y=E$ and $E_Z=0$, which results in
\begin{eqnarray}
\tau_X&=&-E^2  \left(\alpha_{ZX}+\alpha_{ZY}  \right)=-\tau_Y   \nonumber\\
\tau_Z&=&E^2\left(  \alpha_{XX}-\alpha_{YY} \right).
\label{Torque components}
\end{eqnarray}
In case the $a$ axis points along the $X$ direction (Fig. \ref{Simplified figure}a), the torques can be expressed as:
\begin{eqnarray}
-\tau_X&=&\tau_Y\sim  \left( \alpha_{ZX}+\alpha_{ZY}\right)=\alpha_{ba}\sin\psi+\alpha_{ca}\cos\psi\nonumber\\
&+&\alpha_{cb}\cos\left(2\psi\right)+\frac{1}{2}\left(\alpha_{bb}-\alpha_{cc}\right)\sin \left(2\psi\right).
\label{T develop}
\end{eqnarray}
Before application of the second laser pulse, the mean value of the $Z$ component of the permanent dipole moment $\langle\mu_Z\rangle = \langle\mu_b \sin\psi + \mu_c \cos\psi\rangle$ is zero. In order to understand which way this value starts changing just after the second pulse, we consider the quantity $d \langle\mu_Z\rangle/dt$. It is clear that only the torque along the $X$ axis provides a non-vanishing contribution: $d \langle\mu_Z \rangle/dt = \langle\mu_b \cos\psi\ d\psi /dt - \mu_c \sin\psi \ d\psi /dt\rangle$. Considering that after the second impulsive kick $d\psi /dt \sim {\rm const} +\tau_X$, we obtain
\begin{equation}
d \langle\mu_Z\rangle/dt \sim \langle\mu_b\alpha_{ca}\cos^2 \psi - \mu_c\alpha_{ba}\sin^2\psi\rangle  \\
\sim (\mu_b\alpha_{ca}-\mu_c\alpha_{ba}).
\label{avmu}
\end{equation}
In the case when the $a$ axis points against the $X$ direction (Fig. \ref{Simplified figure}b), it can be shown that $\alpha_{ZX}$ and $\alpha_{ZY}$ (as well as $\tau_X,\tau_Y$) change their signs so that the  projection of the  torque along the ${\rm O-S}$ direction remains the same (as illustrated by the black arrows in Fig. \ref{Simplified figure}). As a result, we again arrive to the Eq. \ref{avmu}.

 Therefore, the interaction with the second cross-polarized pulse not only induces the unidirectional rotation of the most polarizable $a$-axis (like in the case of linear  and symmetric molecules \cite{bib22,bib23,Korech,Kenta,JW}), but also initiates molecular rotation {\it about} this axis which results in the {\it orientation} of the average dipole moment along the $Z$ direction. As follows from Eq. \ref{avmu},  this effect requires non-zero values of the off-diagonal elements of the polarizability tensor, which is a typical situation for a generic asymmetric molecule.

\section{Rotational Dynamics}\label{The Theory}

We calculate the ensemble-averaged $Z$-projection of the molecular permanent dipole moment with help of a Monte Carlo simulation. Initially, a large number of sample molecules are randomly distributed in the molecular phase space, i.e. in Euler angles $\phi,\theta,\chi$ (denoting the rotation of the asymmetric rotor principal axes ($a,b,c$) with respect to the laboratory axes ($X,Y,Z$)) and in canonical momenta $P_{\phi},P_{\theta},P_{\chi}$ (see Sec. \ref{Free Space}) according to the canonical thermal distribution function (Sec. \ref{Thermal Ensemble}). The first short laser pulse  modifies this distribution (Sec. \ref{Laser Interaction}), after which every molecule rotates freely (Sec. \ref{Free Space}). These two steps, i.e. laser interaction and free rotation, are repeated for the second laser pulse after a certain time delay. Below we describe in detail every step of our procedure.

\subsection{Free Rotation} \label{Free Space}

The Lagrangian of an asymmetric rotor in free space is given by:
\begin{equation}
L=\frac{1}{2}I_a\Omega_a^2+\frac{1}{2}I_b\Omega_b^2+\frac{1}{2}I_c\Omega_c^2,
\label{Lagrangian}
\end{equation}
where $\Omega_a, \Omega_b$ and $\Omega_c$ are the angular velocity components of the rotor expressed with respect to the body-fixed principal axes $a, b$ and $c$, respectively. They are related to the angular momenta components by:
\begin{eqnarray}
M_a&=&I_a\Omega_a\nonumber\\
M_b&=&I_b\Omega_b\nonumber\\
M_c&=&I_c\Omega_c
 \label{Angular Momenta}.
\end{eqnarray}
Here $I_{a,b,c}$ are the principal moments of inertia, where by convention $I_c \geq I_b \geq I_a$.
The components of the angular velocity along  the $a,b,c$ axes can be expressed by the Euler angles $\phi,\theta$ and $\chi$ and their derivatives \cite{bib35}:
\begin{eqnarray}
\Omega_a&=&-\dot{\phi}\sin\theta\cos\chi+\dot{\theta}\sin\chi \nonumber\\
\Omega_b&=&\dot{\phi}\sin\theta\sin\chi+\dot{\theta}\cos\chi \nonumber\\
\Omega_c&=&\dot{\phi}\cos\theta+\dot{\chi }
 \label{Free Motion}.
\end{eqnarray}

Using Eq. \ref{Free Motion}, the following set of differential equations can be obtained:
\begin{eqnarray}
\dot{\phi}&=&\frac{  -\cos\chi\Omega_a+\sin\chi\Omega_b }{ \sin\theta   } \nonumber\\
\dot{\theta}&=&\sin\chi\Omega_a+\cos\chi\Omega_b\nonumber\\
\dot{\chi}&=&\Omega_c-\cos\theta\frac{ -\cos\chi\Omega_a+\sin\chi\Omega_b  }{ \sin\theta  }
 \label{Differential equation}.
\end{eqnarray}

The relations between $M_a, M_b$ and $M_c$ to the laboratory frame angular momenta $M_X, M_Y$ and $M_Z$ are given by the transformation:

\begin{eqnarray}
    \left(    \begin{matrix}
        M_a \\
        M_b \\
M_c
    \end{matrix}\right)
 &=& \left(\begin{matrix}
       \cos\chi & \sin\chi & 0\\
	 -\sin\chi & \cos\chi& 0\\
	 0 & 0& 1
     \end{matrix}\right)
\left(    \begin{matrix}
       \cos\theta & 0& -\sin\theta\\
	 0 & 1 & 0\\
	 \sin\theta & 0& \cos\theta
    \end{matrix}\right)
\nonumber\\	
&\times&  \left(    \begin{matrix}
       \cos\phi & \sin\phi & 0\\
	 -\sin\phi & \cos\phi& 0\\
	 0 & 0& 1
    \end{matrix}\right)
\left(    \begin{matrix}
        M_X \\
        M_Y \\
M_Z
    \end{matrix}\right)
\label{Transformation}.
\end{eqnarray}

It is important to note that $M_X,M_Y,M_Z$ are constant in  the course of a free rotation, while  $\Omega_a,\Omega_b,\Omega_c$ are related to them according to Eqs. \ref{Angular Momenta} and \ref{Transformation}.
Considering Eqs.  \ref{Differential equation}, \ref{Angular Momenta} and \ref{Transformation}, the Euler angles $\phi,\theta,\chi$ are numerically calculated at any time  by solving the set of coupled differential equations with the initial conditions, i.e. the initial Euler angles and the initial angular momenta.  The latter can be derived from the initial canonical momenta  given by

\begin{eqnarray}
P_{\phi}&=&\frac{\partial L}{\partial \dot{\phi}}=-\sin\theta\cos\chi M_a+\sin\theta\sin\chi M_b +\cos\theta M_c\nonumber\\
P_{\theta}&=&\frac{\partial L}{\partial \dot{\theta}} =\sin\chi M_a +\cos\chi M_b\nonumber\\
P_{\chi}&=&\frac{\partial L}{\partial \dot{\chi}}=M_c
\label{Canonical Momenta}.
\end{eqnarray}

\subsection{Interaction with a Laser Pulse } \label{Laser Interaction}

The interaction potential describing coupling of a laser field with the molecular polarizability is given by \cite{Craig,bib39}:
\begin{equation}
V(t)=-\frac{1}{4}\Sigma_{\rho\rho '}\epsilon_{\rho}(t)\alpha_{\rho \rho '}\epsilon_{\rho '} ^{\**}(t),
\label{Potential}
\end{equation}
where $\epsilon$ is the envelope of the laser electric field, $\alpha$ is the polarizability tensor, and $\rho,\rho '$ correspond to the laboratory axes $X,Y,Z$.
The relation between the polarizability components in the space fixed axes and the polarizability components in the body fixed frame is:

\begin{equation}
\alpha_{\rho \rho '}
= \left(\begin{matrix}
     \braket{  \rho |  a } & \braket{\rho   | b} & \braket{\rho   |  c } \\
     \end{matrix}\right)
\left(    \begin{matrix}
       \alpha_{aa} & \alpha_{ab}& \alpha_{ac}\\
	 \alpha_{ba} & \alpha_{bb}& \alpha_{bc}\\
	\alpha_{ca} & \alpha_{cb}& \alpha_{cc}
    \end{matrix}\right)	
 \left(    \begin{matrix}
      \braket{ a| \rho '}\\
 \braket{b |\rho '} \\
\braket{ c| \rho '}
    \end{matrix}\right),
\label{Polarizability}
\end{equation}
 where $\braket{k|\rho}$ are the direction cosines.
If the pulse is short compared to the typical period of molecular rotation, it may be considered as a delta pulse.
Considering the  Euler-Lagrange equations:
\begin{eqnarray}
 \frac{dP_{\phi}}{dt}-\frac{\partial L}{\partial \phi} &=& 0\nonumber\\
 \frac{dP_{\theta}}{dt}-\frac{\partial L}{\partial \theta} &=& 0\nonumber\\
 \frac{dP_{\chi}}{dt}-\frac{\partial L}{\partial \chi} &=& 0,
\label{Euler Lagrange}
\end{eqnarray}
(here $L$ is the Lagrangian of the asymmetric rotor in the field), in the impulsive approximation, one finds the effect of a short laser pulse on the canonical momenta:
\begin{eqnarray}
P_{\phi}=P_{\phi}(0)-\int{dt\frac{\partial V}{\partial \phi}}\nonumber\\
P_{\theta}=P_{\theta}(0)-\int{dt\frac{\partial V}{\partial \theta}}\nonumber\\
P_{\chi}=P_{\chi}(0)-\int{dt\frac{\partial V}{\partial \chi}}.
\label{Canonical Momenta Change}
\end{eqnarray}
Here $P_{\phi}(0), P_{\theta}(0),P_{\chi}(0)$ are the values of the canonical momenta just before the laser pulse is applied, and integration is performed over the duration of the short laser pulse.
The derivatives of the interaction potential in the rhs of Eq. \ref{Canonical Momenta Change}  involve derivatives of  different components of the polarizability tensor $\alpha_{\rho \rho '}$. According to Eq. \ref{Polarizability}, each derivative includes two terms:

\begin{eqnarray}
\frac{\partial \alpha_{\rho \rho '}}{\partial \delta}
&=& \left(\begin{matrix}
     \frac{\partial }{\partial \delta}\braket{  \rho |a} &  \frac{\partial }{\partial \delta}\braket{  \rho |b } &  \frac{\partial }{\partial \delta}\braket{  \rho  |c} \\
     \end{matrix}\right)\nonumber\\
&\times&
\left(    \begin{matrix}
       \alpha_{aa} & \alpha_{ab}& \alpha_{ac}\\
	 \alpha_{ba} & \alpha_{bb}& \alpha_{bc}\\
	\alpha_{ca} & \alpha_{cb}& \alpha_{cc}
    \end{matrix}\right)	
 \left(    \begin{matrix}
      \braket{ a| \rho '}\\
 \braket{b |\rho '} \\
\braket{ c| \rho '}
    \end{matrix}\right)\nonumber\\
&+& \left(\begin{matrix}
     \braket{  \rho   |a} & \braket{ \rho  |b} & \braket{ \rho   |c} \\
     \end{matrix}\right)
\left(    \begin{matrix}
       \alpha_{aa} & \alpha_{ab}& \alpha_{ac}\\
	 \alpha_{ba} & \alpha_{bb}& \alpha_{bc}\\
	\alpha_{ca} & \alpha_{cb}& \alpha_{cc}
    \end{matrix}\right)\nonumber\\
&\times&	
 \left(    \begin{matrix}
       \frac{\partial}{\partial \delta} \braket{ a| \rho '}\\
\frac{\partial}{\partial \delta} \braket{ b| \rho '}\\
\frac{\partial}{\partial \delta} \braket{ c| \rho '}
    \end{matrix}\right),
\label{Polarizability Derivative}
\end{eqnarray}

where $\delta$ is either $\phi,\theta$ or $\chi$.

\subsection{Thermal Ensemble} \label{Thermal Ensemble}
Considering a canonical ensemble, the initial thermal distribution function for the rotating molecules has the following form in the dimensionless variables (see, e.g. \cite{bla}):
\begin{equation}
f(\phi,\theta,\chi,\bar{\bar{P_{\phi}}},\bar{P_{\theta}},\bar{P_{\chi}})=\sin\theta\frac{\exp{\left[-\frac{1}{2}\left(\bar{\bar{P_{\phi}}}^2+ \bar{P_{\theta}}^2  +\bar{P_{\chi}}^2   \right)\right]}}{\sqrt{\pi\left(8\pi^2 \right)^3}  },
\label{canonical distribution}
\end{equation}
where
\begin{eqnarray}
\bar{P_{\theta}}&=&\left( P_{\theta}+\frac{\bar{P_{\phi}}}{\sin\theta}\frac{\cos\chi\sin\chi\left(\frac{1}{I_b}-\frac{1}{I_a}\right)}{\frac{\sin^2\chi}{I_a}+\frac{\cos^2\chi}{I_b}}   \right)\nonumber\\
&\times&\sqrt{\frac{\frac{\sin^2\chi}{I_a}+\frac{\cos^2\chi}{I_b}}{k_BT}} \nonumber \\
\bar{P_{\phi}}&=&P_{\phi}-\cos\theta P_{\chi}\nonumber \\
\bar{\bar{P_{\phi}}}&=&\frac{\bar{P_{\phi}}}{\sin\theta}\sqrt{\frac{1}{k_B T I_a I_b \left(   \frac{\sin^2\chi}{I_a}  +\frac{\cos^2\chi}{I_b}  \right)}}\nonumber\\
\bar{P_{\chi}}&=&\frac{P_{\chi}}{\sqrt{I_c k_B T}}.
\label{change variables}
\end{eqnarray}
Here $T$ is the  temperature of the gas, and $k_B$ is the Boltzmann constant.

The initial  ensemble of molecules with different Euler angles and canonical momenta was generated via a Monte Carlo procedure according to the distribution of Eq. \ref{canonical distribution}. The time-dependent $Z$-projection of the molecular dipole moment ($\mu_Z$) was  calculated for each molecule in the ensemble, and finally, the result was averaged over the initial distribution:

\begin{eqnarray}
\braket{\mu_Z}=\int_0^{2\pi} d\phi \int_0^{2\pi} d\chi \int_0^{\pi} d\theta \int_{-\infty}^{\infty} d\bar{P_{\chi}} \nonumber\\
\int_{-\infty}^{\infty} d\bar{\bar{P_{\phi}}}    \int_{-\infty}^{\infty}  d\bar{P_{\theta}}  f(\phi,\theta,\chi,\bar{\bar{P_{\phi}}},\bar{P_{\theta}},\bar{P_{\chi}})\mu_Z.
\label{Thermal}
\end{eqnarray}

\section{Results and Discussion} \label{Results}
As in Sec. \ref{Simplified Classical Analysis},  we consider the ${\rm HSOH}$ molecule as an example.
The molecule dipole moments (in Debye) are \cite{bib33} $\mu_a=0.053$, $\mu_b=0.744$ and $\mu_c=1.399$. The polarizability components (in atomic units) are \cite{bib33} $\alpha_{aa}=31.85$, $\alpha_{bb}=26.20$, $\alpha_{cc}=26.51$, $\alpha_{ab}=-0.94$, $\alpha_{ac}=-0.84$ and $\alpha_{bc}=0.07$. The rotational constants (in MHz) are \cite{bib34}: $A=202136$, $B=15279$ and $C=14840$. The laser pulses considered in this work have peak intensity of $5.3  \times  10^{13} \ {\rm W/cm^2}$ and $100 \ {\rm{fs}}$ full width at half maximum (FWHM) duration.

\begin{figure}[htb]
\begin{center}
\includegraphics[width=8cm]{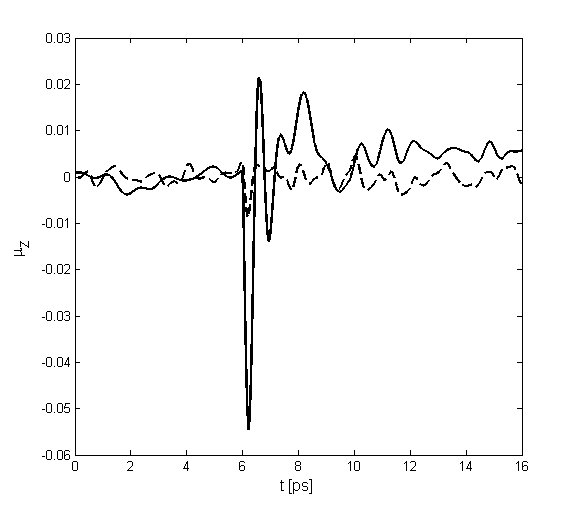}
\end{center}
\caption{Time evolution of the ensemble averaged $\mu_Z$ component of the permanent dipole moment is plotted for an enantiomer of ${\rm HSOH}$ molecule subject to a double-pulse excitation at $T=0K$ (solid line) and $50K$ (dashed line). The details about the laser pulses are given in the text.}
\label{mu Z vs time delay figure}
\end{figure}

Following the classical procedure presented in Sec. \ref{The Theory}, we calculate the average $Z$-projection  of the permanent dipole moment, $\langle\mu_Z\rangle$ for  ${\rm HSOH}$ molecules subject to a pair of time delayed ($6ps$ time delay) linearly polarized laser pulses. The first pulse is polarized along the $X$ direction and polarization vector of the second pulse is confined to the $XY$ plane and constitutes $45^{\circ}$ with the $X$ direction. Fig.  \ref{mu Z vs time delay figure} shows the double-pulse induced average dipole moment, $\langle\mu_Z\rangle$  both at $T=0K$ and $T=50K$ (more than $300000$ sample molecules in the molecular phase space were used in the Monte Carlo simulations in order to prepare Fig. \ref{mu Z vs time delay figure}).

\begin{figure}[htb]
\begin{center}
\includegraphics[width=8cm]{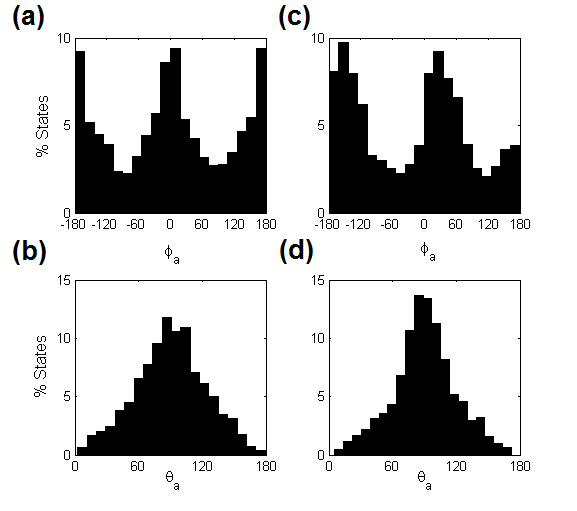}
\end{center}
\caption{Distributions of  spherical polar angles  $\phi_a$ (a) and $\theta_a$ (b)  of the $a$ axis just before the second pulse is applied ($T=0K$). As seen, the $a$ axis is preferentially  aligned along the $\pm X$ directions at this time moment. Panels (c) and (d) show the distributions of the spherical polar angles $\phi_a$ and $\theta_a$, respectively,  at the moment of the first peak of $|\langle\mu_Z\rangle|$ after the second pulse (see Fig. \ref{mu Z vs time delay figure}).  }
\label{a distribution}
\end{figure}

  According to Fig. \ref{mu Z vs time delay figure}, the $\langle\mu_Z\rangle$ value is zero after the first pulse (which  obviously follows from the symmetry considerations). However, $\langle\mu_Z\rangle$ takes finite non-zero values after the second pulse (as was predicted by Eq. \ref{avmu}).
  Our Monte Carlo simulations directly support the qualitative explanation of the orientation mechanism presented in Sec. \ref{Simplified Classical Analysis}. We define the spherical polar angles of the $a$ axis with respect to the $X, Y, Z$ axes as $\phi_a$ , $\theta_a$ and plot their distributions in Fig. \ref{a distribution} ($T=0K$). Figs. \ref{a distribution}a and \ref{a distribution}b show the distributions just before the second pulse. Figs. \ref{a distribution}c and \ref{a distribution}d correspond to $\phi_a$ and $\theta_a$ at the moment of the first peak of $|\langle\mu_Z\rangle |$ in Fig. \ref{mu Z vs time delay figure} after the second pulse.
As follows from  Figs. \ref{a distribution}a and \ref{a distribution}b, just before the second pulse  the $a$ axis is preferentially aligned along the $\pm X$ direction (the distribution of $\phi_a$ has maxima around $0^{\circ}$ and $180^{\circ}$,  and distribution of $\theta_a$ is maximal around $90^{\circ}$). The role of the first pulse  is therefore to induce {\it alignment} of the molecular $a$ axis along the $X$ direction.
As was shown in the past \cite{bib22,bib23,bib24}, the $Z$-component of the torque applied by the second pulse to the aligned molecules  should cause their unidirectional rotation about $Z$ axis.  As can be seen from  Figs. \ref{a distribution}c and \ref{a distribution}d, the most polarizable axis of the ${\rm HSOH}$ molecules ($a$ axis) is indeed shifted toward the polarization direction of the second pulse in the $XY$ plane at the moment of the $\langle\mu_Z\rangle$ peak (see Fig. \ref{mu Z vs time delay figure}). A sharp dipole signal seen in Fig. \ref{mu Z vs time delay figure} after the second pulse  appears due to the component of the torque {\it parallel} to the alignment direction.

\section{Differentiation of Enantiomers} \label{Differentiation of Enantiomers}
\begin{figure}[htb]
\begin{center}
\includegraphics[width=8cm]{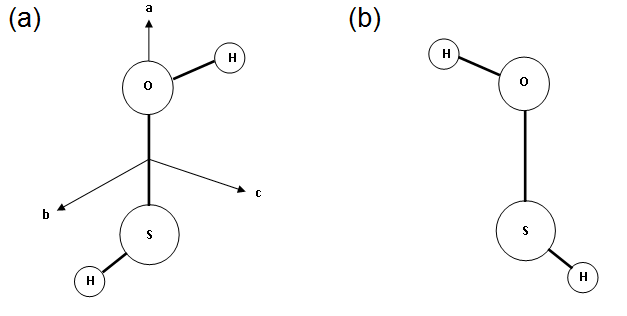}
\end{center}
\caption{(a) and (b) show two enantiomers of the ${\rm HSOH}$ molecule.}
\label{Definitions figure}
\end{figure}

As it was already mentioned above,  generic asymmetric molecules are essentially chiral in nature. Therefore, an immediate possible application of the described laser induced orientation is enantiomer differentiation of chiral molecules \cite{bib33}. Considering the example of the ${\rm HSOH}$ molecule, two different enantiomers of it are shown in Figs. \ref{Definitions figure} (a) and (b).
The molecule's dipole moments and polarizability components have already been given in Sec. \ref{Results}, however $\mu_c$ has opposite signs for the two enantiomers, as well as the off-diagonal polarizability components $\alpha_{ac}$ and $\alpha_{bc}$ \cite{bib33}.

As discussed above, a delayed pair of crossed polarized pulses induces unidirectional rotation of the most polarizable molecular axis, and the sense of this rotation is the same for both enantiomers. In addition, the second laser pulse induces a torque along the aligned molecular axis and orients the molecules. Because of the opposite signs of the off-diagonal polarizability components for left- and right-handed enantiomers, the induced torque has different signs for different enantiomers, which leads to the counter-rotation of their permanent dipole moments. This conclusion is supported by Eq. \ref{avmu}, that describes the rate of change of the mean value of the vertical (Z) component of dipole moment just after the second pulse. The rhs of Eq. \ref{avmu} has opposite signs for different enantiomers ($\alpha_{ca}$ and $\mu_c$ change sign between different enantiomers).

\begin{figure}[htb]
\begin{center}
\includegraphics[width=8cm]{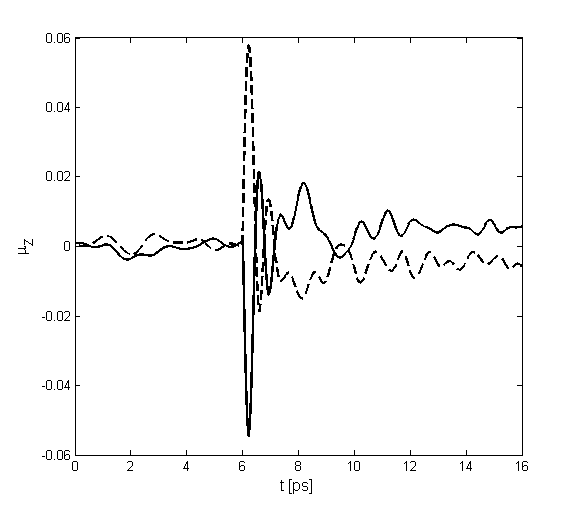}
\end{center}
\caption{Time evolution of the ensemble averaged $\mu_Z$ component of the permanent dipole moment for two enantiomers of  ${\rm HSOH}$ molecule subject to a double-pulse excitation ($0K$).  The same conditions as in Fig. \ref{mu Z vs time delay figure} are maintained, such as pulses delay and intensities.}
\label{mu Z figure}
\end{figure}

Fig. \ref{mu Z figure} shows results of a direct classical calculation of the time-dependence of  ensemble-averaged  $Z$-projection of the molecular permanent dipole moment, $\langle\mu_Z\rangle$ for the two enantiomers of ${\rm HSOH}$ molecule at at $T=0K$.

One can clearly observe in Fig. \ref{mu Z figure} the out-of-phase time evolution of the dipole signals produced by the two enantiomers (compare with \cite{bib33}). This allows differentiation of enantiomers, detection of enantiomeric excess  and other possible manipulations in a racemic mixture. The classical nature of the differentiation mechanism makes it robust and potentially operational in dissipative environment.

\section{Conclusions} \label{conclusions}

A single linearly polarized laser pulse defines an axis (a non-oriented direction) in space. It is able to align (but not orient) anisotropically polarizable molecules  along this axis. However, two delayed pulses with crossed polarizations define an oriented vector (unless the polarizations are exactly orthogonal). It is perpendicular to the plane spanned by the two polarization axes, and directed along the rotation vector needed to bring the first axis to the direction of the second one. It was shown in the past that such a pair of pulses is capable of orienting the angular momentum of linear and symmetric molecules along the above defined direction in space \cite{bib22,bib23,bib24}. In this work we extended the problem to asymmetric molecules, and showed that such an excitation  may orient the molecules themselves via  interaction of the laser field with the induced polarization. We described the  mechanism causing this orientation, and showed that it is classical in nature.  The first laser pulse causes transient alignment of the most polarizable molecular axis. For a generic asymmetric molecule, the second pulse not only initiates the unidirectional rotation of this axis, but also induces a mechanical torque {\it along} it.  This results in the unidirectional  molecular rotation {\it about} the aligned axis. As the direction of the molecular dipole moment in asymmetric molecules is generally different from the direction of the most polarizable axis, the above mentioned torque causes a transient orientation of the ensemble-averaged dipole moment along (or against) the directed vector defined by the crossed polarizations of the pulses. The orientation mechanism described in this paper provides a novel valuable addition to  the already existing toolbox of laser methods for molecular orientation, which use various approaches for introducing  directional asymmetry in the light-molecule interaction. In our case, orientation of asymmetric molecules is achieved by a chiral skewing of the polarization of short non-resonant laser pulses. For two chiral enantiomers, their dipoles tend to be oriented in opposite directions due to the opposite signs of their off-diagonal polarizability components and the components of the  dipole moments in the molecular frame.  This results in the out-of-phase oscillations of the enantiomers' dipole moments \cite{bib33}, which may serve for the chiral analysis purposes, say by phase-sensitive measurements of the laser-induced emission from the irradiated gas samples \cite{m7,m10}. Quantum revivals of the laser-induced dipole orientation signals may be useful for extracting information on molecular rotational dynamics and dissipation effects \cite{Harde,Fleischer1,Fleischer2,Faucher2016}, and for studying collective phenomena in molecular emission \cite{FleischerDecay}. In addition to the double pulse scheme considered in this paper, other implementations of the chiral polarization skewing may be useful, including  optical centrifuge  \cite{Corkum,Villeneuve,Mullin,Korobenko2014}, chiral pulse trains \cite{Valerytrain}, or continuous polarization twisting by laser field shaping \cite{Karras2015}). Deflecting pre-oriented chiral molecules by inhomogeneous laser, static and magnetic fields \cite{bib40,bib41} may be promising for enantiomer separation.

\section{Acknowledgements} \label{Acknowledgements}

 Support by the Israel Science Foundation  (Grant No. 746/15)  is highly appreciated.  We thank V. Milner and T. Momose for valuable discussions.  I.A. acknowledges support as the Patricia Elman Bildner Professorial Chair. This research was made possible in part by the historic generosity of the Harold Perlman Family.

\bibliographystyle{phaip}

\end{document}